\documentclass[letterpaper]{article}
\usepackage[draft]{aaai2027}                   % AAAI 官方宏包
\usepackage{times}                      % 规范 Times 字体
\usepackage{helvet}                     % 规范无衬线字体
\usepackage{courier}                    % 等宽字体
\usepackage[hyphens]{url}               % URL 支持
\usepackage{graphicx}                   % 插图宏包
\usepackage{amsmath,amssymb}            % 数学公式
\usepackage{booktabs}                   % 学术三线表
\usepackage{enumitem}                   % 列表定制（支持 leftmargin 等可选参数）
\usepackage{caption}
\title{Structured Interaction, Visual Localization, and Robust Execution for Complex Web Tasks: A Technical Report on the WebRetriever Challenge}
\author{Shaohui Li, Ziqi Zhang, Bing Li}
\affiliations{Beijing Key Laboratory of Multimodal Super-Intelligent Security \& People AI}

\begin{document}

\maketitle

\begin{abstract}
This report presents the web agent system developed for the WebRetriever Challenge. The system follows a structured-interaction-first strategy, using semantic webpage information for routine browser operations and invoking visual perception only when structured representations are insufficient. Three key designs are introduced: grid-assisted visual localization for difficult-to-access controls, hierarchical context management for reducing redundant page and interaction history, and fault-aware execution mechanisms for stable multi-browser task processing. The system achieved a pass rate of up to 79\% in local evaluation on Protocol~1. In the official Protocol~3 competition, it achieved a 59\% pass rate with eight concurrent browser workers and ranked first overall, winning the WebRetriever Challenge. 
\end{abstract}

\section{Introduction}

With the rapid progress of large language models (LLMs) in reasoning, planning, and tool use, web automation is evolving from rule-based scripts toward autonomous agents driven by natural-language instructions. 
Unlike conventional automation systems, web agents must understand the current page state, autonomously select interaction targets and operation types, and continuously adjust subsequent actions according to feedback from the website. 
These tasks combine webpage understanding, multi-step decision making, and browser control.

The WebRetriever Challenge \cite{dong2026webretriever} requires an agent to operate remote browsers provided by the organizers and complete information retrieval, filtering, sorting, and content extraction tasks in realistic or near-realistic web environments. 
Task completion may require applying website constraints and inspecting the resulting page, in addition to retrieving relevant information.

During system development, we identified three major challenges. 
First, structured representations such as accessibility trees and the Document Object Model (DOM) can effectively describe buttons, input fields, and textual content, but provide incomplete information for charts, images, colors, selection states, and certain non-standard controls. 
Second, as the number of interaction steps increases, page snapshots, script outputs, and historical actions accumulate rapidly. 
The resulting redundancy consumes the model context and may obscure information that is critical for subsequent decisions. 
Third, the official competition requires the continuous execution of a large number of tasks while maintaining multiple remote browser instances. 
Failures in browser connections, MCP services, page loading, or tool execution may not only affect the current task but also propagate to subsequent tasks assigned to the same worker.

To address these challenges, we build a web agent system based on AgentScope~\cite{agentscope} and Playwright MCP~\cite{playwrightmcp}, with additional components for visual localization, long-horizon context management, and multi-task scheduling with fault recovery. 
The overall system follows the principle of \emph{structured interaction first, visual perception on demand}. 
These components address complementary requirements: accessing visually represented information, retaining relevant task state within a bounded context, and isolating runtime failures across tasks.

\section{System Architecture and Task Execution}

\begin{figure}[t]
\centering
\includegraphics[width=\columnwidth]{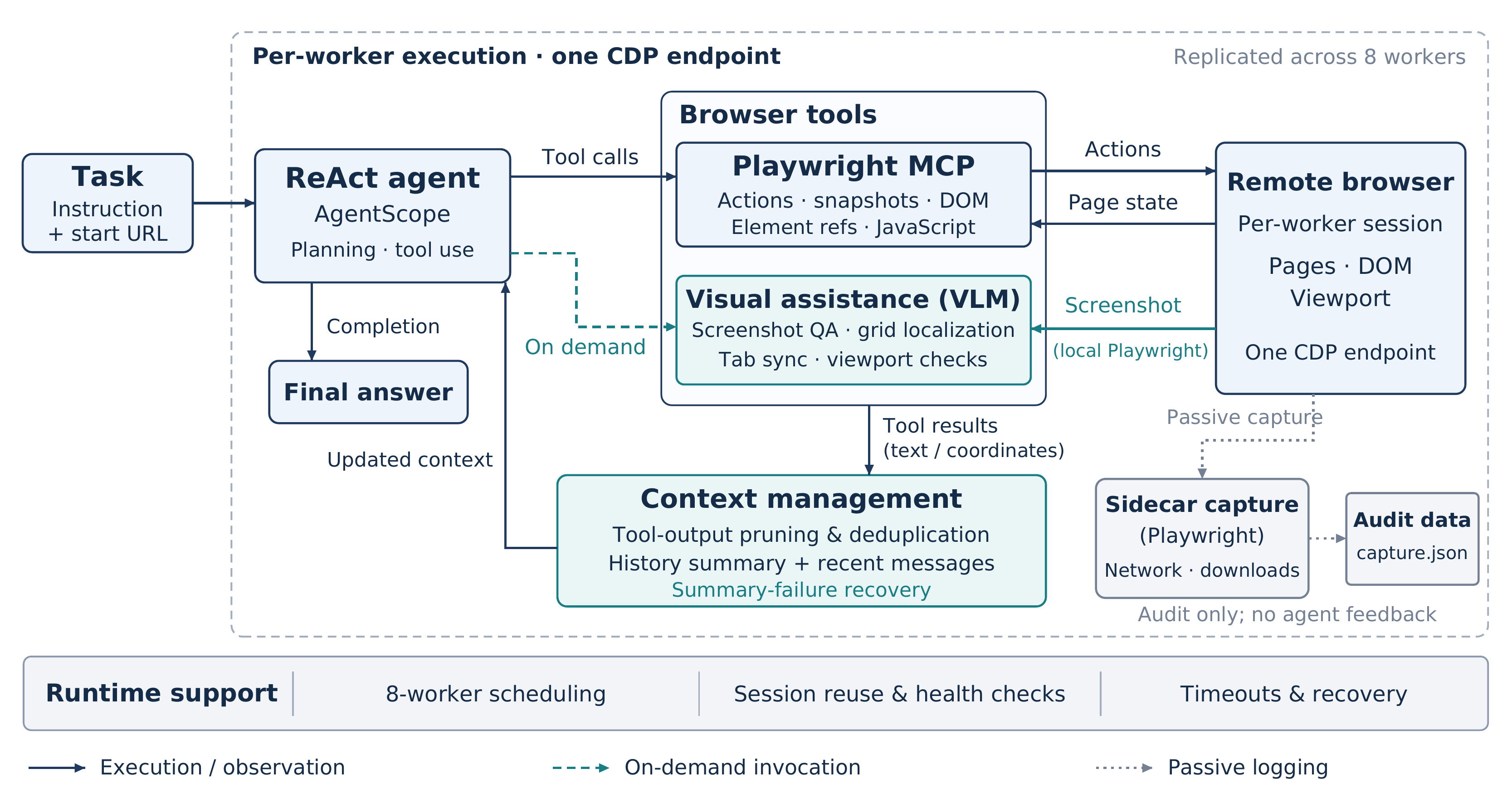}
\caption{Architecture of the WebRetriever system. Each task is executed by an independent main agent that interacts with the browser via CDP.}
\label{fig:webretriever_arch}
\end{figure}

Each task is executed by an independent main agent. 
The agent receives a natural-language task description, the starting website, and a task identifier, and connects to the remote browser through the Chrome DevTools Protocol (CDP). 
During execution, the main model selects the next action according to the task instruction, previous operation history, and current page state. 
The corresponding tool is then invoked, and its output is appended to the context for the next decision step. 
This process continues until the task is completed or a termination condition is reached.

The execution loop is implemented using the ReAct paradigm~\cite{react} provided by AgentScope~\cite{agentscope}, which organizes model invocation, tool execution, and observation feedback. 
Browser interaction is primarily supported by Playwright MCP~\cite{playwrightmcp}, including page navigation, page snapshots, clicking, text input, form filling, tab management, and JavaScript execution. 
Among these tools, \texttt{browser\_snapshot} converts the page accessibility tree into textual representations containing element names, roles, and references for interactive components. 
The model can therefore select an element according to its semantics and directly invoke a click or input operation through the corresponding reference, without predicting pixel coordinates.

For more fine-grained inspection of page states, the system also provides a JavaScript execution interface that allows the agent to query DOM elements, attributes, and text content. 
The system prompt explicitly requires the model to use URLs, links, and interaction targets that are either supplied by the user or actually observed from the current website, rather than guessing unseen URL paths or page elements. 
Similarly, JavaScript tools are used only to inspect or manipulate the actual webpage and are not allowed to fabricate task results independently of the website.

WebVoyager~\cite{webvoyager} demonstrates the use of multimodal models for web interaction. Our system keeps browser control with the main agent and provides screenshot analysis and coordinate localization as auxiliary tools, invoked when structured information is insufficient.

\section{Visual Assistance for Page Understanding and Localization}

When the available structured page information is insufficient for task reasoning, the main agent can invoke a screenshot analysis tool, which captures the current browser viewport and sends it to the vision-language model together with a specific visual question. 
For example, when a task requires interpreting a bar chart, the agent may ask which category corresponds to the tallest bar in the current viewport. 
The tool also returns the current page scroll position, enabling the main agent to determine where the screenshot lies within the webpage and whether additional scrolling or page inspection is required. 
Since screenshot analysis only covers the current viewport, its output is not treated as a complete description of the entire page.

For controls that are visible in a screenshot but cannot be reliably localized through element references or DOM information, the system provides a visual coordinate localization tool. 
The tool overlays a regular grid with \textbf{16 columns and 12 rows} on the screenshot and adopts a normalized coordinate system ranging from 0 to 1000 along both axes.

\begin{figure}[t]
    \centering
    \includegraphics[width=\columnwidth]{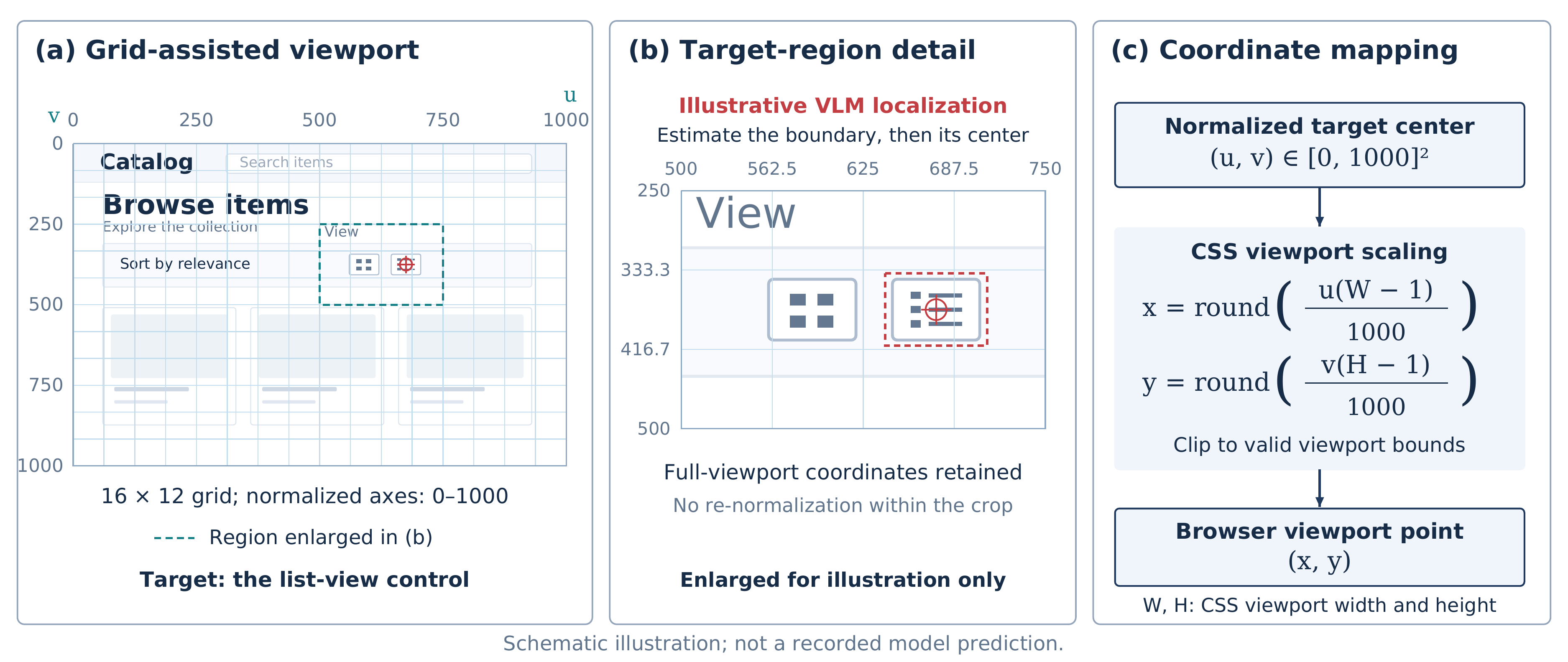}
    \caption{Illustration of the grid-based visual coordinate localization. The viewport is divided into a $16 \times 12$ grid; each cell covers a fixed region, and the coordinate system is normalized to $[0,1000] \times [0,1000]$.}
    \label{fig:grid_localization}
\end{figure} 
Instead of directly predicting pixel coordinates, the vision-language model can first identify the approximate grid region containing the target and then estimate a more precise normalized position. 
The localization prompt further instructs the model to first describe the approximate boundary of the target control and then output a central point within that region, reducing the likelihood of confusing the target with adjacent text, icons, or enclosing containers.

Let $(u,v)\in[0,1000]^2$ be the normalized position predicted by the visual model, and let $W$ and $H$ denote the browser CSS viewport dimensions. 
The corresponding browser coordinates are computed as
\begin{align*}
x &=
\operatorname{round}
\left(
\frac{u}{1000}(W-1)
\right),\\
y &=
\operatorname{round}
\left(
\frac{v}{1000}(H-1)
\right).
\end{align*}
The resulting coordinates are further clipped to the valid viewport range. 
The tool layer thus maps resolution-independent model outputs to the actual browser viewport. 
The localization tool also validates the returned fields, numerical types, and coordinate ranges. 
If the visual model cannot identify the target or returns an unparsable result, the tool outputs an empty coordinate and leaves the main agent to decide whether to inspect the page again or adopt an alternative interaction strategy. 
The localization tool itself does not perform the click; all mouse operations remain under the unified browser control interface.

Browser operations are executed through Playwright MCP, whereas screenshots are acquired through a local Playwright client connected to the same remote browser. 
To ensure consistency between the page being operated on and the page being captured, the synchronization module reads the current page information returned by MCP, identifies the corresponding tab, and activates it before screenshot capture. 
The CSS viewport dimensions are also measured before and after localization. 
If the viewport size changes, the screenshot is reacquired to avoid converting coordinates using inconsistent image and viewport dimensions. 
This procedure reduces errors caused by window-size changes, although it cannot fully eliminate coordinate drift due to animations, asynchronous loading, or page reflow. 

\section{Context Management for Long-Horizon Tasks}

Complete page snapshots, large JavaScript outputs, and long execution trajectories can rapidly exhaust the agent context. Our system controls this growth at two complementary levels: individual tool outputs and long-term interaction history.

The accessibility tree returned by \texttt{browser\_snapshot} often contains navigation bars, repeated buttons, long lists, and other information that is only weakly related to the current task. 
If a snapshot is smaller than \textbf{48,000 bytes}, it is retained in full. 
Otherwise, the system performs structure-aware truncation according to page roles and hierarchical relationships. 
Regions such as \texttt{main}, \texttt{article}, and \texttt{dialog} are preferentially retained, together with major headings, tabs, forms, tables, and interactive controls. 
Additional page content is added only when the remaining budget permits. 
Original element references are preserved whenever possible, so that the agent can continue to interact with retained elements even after part of the textual page representation has been omitted. 
If no reliable main region can be identified, the system falls back to prioritizing headings and actionable elements across the entire snapshot. 
This process is rule-based and does not require additional model calls.

JavaScript queries can similarly generate substantial redundancy, particularly when text is simultaneously extracted from parent containers and their child elements. 
For large outputs from \texttt{browser\_evaluate}, the system first attempts to parse the result as JSON, identifies repeated long strings, and replaces redundant occurrences with omission markers while shortening excessively long field values. 
The original field and entry structure is preserved as much as possible. 
The current budget for JavaScript results is \textbf{24,000 bytes}, and the threshold for processing long strings is \textbf{480 bytes}. 
After these specialized compression procedures, the results are further subject to AgentScope's generic tool-output length limits before being inserted into the agent context.

Even after individual tool outputs are controlled, the interaction history continues to grow throughout a long ReAct trajectory. 
The context budget of the main agent is set to \textbf{280,000 tokens}. 
When approximately \textbf{75\%} of this capacity is reached, historical compression is triggered. 
Older interactions are summarized by the model, while approximately the most recent \textbf{10\%} of the original messages are retained.
% Author check: confirm whether the 10 percent retention setting is measured by message count or token count. 
The generated summary is limited to \textbf{6000 characters} and is instructed to preserve the task objective, important completed operations, current page state, key information already obtained, and remaining steps. 
This strategy maintains a compact representation of long-term progress while allowing the model to directly inspect recent tool calls and webpage feedback.

Summary generation itself may fail because of model-service errors or malformed outputs. 
Instead of terminating the entire task in such cases, the recovery module preserves as many recent original messages as permitted by the remaining budget and reinserts the original task together with an instruction to continue execution. 
When necessary, the agent is instructed to inspect the current page again before making the next decision. 
This fallback prioritizes continued execution, although earlier details may be lost and need to be reacquired from the website.

\section{Multi-Task Scheduling and Fault Recovery}

Concurrent execution introduces browser disconnections, tool timeouts, and unresponsive workers in addition to task-level reasoning errors. We distinguish failures by their scope and apply corresponding recovery strategies to limit their impact on subsequent tasks.

The system launches one independent worker process for each CDP browser endpoint. 
During the official competition, \textbf{eight workers} were used to control eight remote browsers concurrently. 
Each worker retrieves an unassigned task from a shared task table. 
A short lock is used only during task assignment to prevent duplicate allocation and is immediately released afterward, so long-running model inference and browser operations do not block other workers from acquiring tasks. 
Tasks are executed sequentially within each worker, while different workers operate in parallel.

To reduce repeated initialization overhead, each worker reuses its local Playwright client, MCP process, and remote browser \texttt{BrowserContext} across tasks. 
Agent states and message histories are reinitialized for every task, ensuring that the model context from one task does not leak into the next. 
The browser session, however, remains active so that necessary website states can be preserved. 
Before receiving a new task, a worker checks both the local Playwright connection and actual browser accessibility through MCP rather than relying solely on the existence of local connection objects. 
After a task is completed, redundant tabs and task-specific handlers are removed while at least one usable page is preserved. 
If the connection becomes invalid, the local client and MCP session are reconstructed without proactively closing the remote browser supplied by the organizers.

For ordinary page-operation or tool failures, the error is first returned to the main model, which may decide to retry, wait for the page, reacquire page information, or use an alternative tool. 
The timeout for a single tool invocation is set to \textbf{180 seconds}. 
The first two consecutive timeouts are returned to the agent as error observations, allowing it to adapt its behavior. 
A third consecutive timeout terminates the current task to prevent the worker from remaining indefinitely blocked by an unresponsive tool. 
The system prompt also limits repeated execution of the same failed action.

If the failure originates from the browser or MCP connection, reconnection is attempted a limited number of times before the task begins. 
When a connection is lost during task execution, the current task is recorded as failed and the runtime environment is recovered before processing subsequent tasks. 
A failure to load an individual website is still treated as a page-level problem and does not immediately trigger reconstruction of the entire browser environment. 
If MCP initialization repeatedly times out, the corresponding worker is marked unavailable and stops accepting new tasks.

More severe worker-level failures are handled by the parent process, which monitors periodic heartbeats and the current execution stage reported by each worker. 
If a worker remains in a connection, navigation, or cleanup stage beyond a predefined waiting threshold, the parent process terminates and restarts it. 
The total execution time for a single task is limited to \textbf{one hour}, and each worker may be automatically restarted at most \textbf{twice}. 
Tasks that have already been assigned are not automatically returned to the shared task queue after a restart. 
Instead, the parent process first checks existing execution results, completes missing result files if necessary, and then continues scheduling tasks that have not yet been assigned. 
This policy prioritizes processing the remaining tasks while retaining the recorded outcome of each assigned task.

\section{Task Completion Judgment and Execution Logging}

Before producing the final answer, the system prompt instructs the main model to verify whether each requirement in the task has been supported by actual webpage evidence. 
For example, if a task requires filtering and sorting before returning an answer, the corresponding conditions should be applied on the actual website and the final page should be inspected, rather than inferring the answer from the initial page. 
If the available evidence is insufficient, the agent is expected to continue inspecting the webpage or performing the required operations. 
If the task still cannot be completed, the final answer should explicitly indicate the incomplete portion.

The current system does not employ a separate task-success verifier. 
Termination is mainly determined by the main model according to the current context. 
Consequently, a normal program return does not necessarily imply that the task is correct according to the official evaluation criteria. 
In the output records, \texttt{SUCCESS} indicates that the agent terminated normally and produced a final response, whereas \texttt{FAIL} indicates an execution exception. 
Actual task success is ultimately determined by the competition evaluator based on the final answer and execution evidence.

For organizer-side verification, we additionally implement a passive Playwright-based logging module that records webpage requests, download events, page URLs, tool-call trajectories, and related screenshots. 
Network records contain the URLs, request methods, parameters, and originating pages of XHR and Fetch requests, while download records contain filenames and completion status. 
Full network response bodies are not stored. 
During task execution, these events are written to a temporary SQLite database and exported to \texttt{capture.json} after the task finishes, avoiding the need to keep all records in memory.
The logging module is decoupled from agent decision making: captured information is not supplied to the main model or used for action selection or completion judgment. 
Each task ultimately produces files including \texttt{result.json}, the execution trajectory, and \texttt{capture.json}, which collectively record the final response, browser interactions, and runtime status.

\section{Implementation Configuration and Experimental Results}

The system is implemented in Python~3.11 using AgentScope~2.0.6, Playwright~1.62.0, and Playwright MCP~0.0.79. Both the main model and the vision-language model use Qwen3.7-Plus. The model temperature is set to 0.2, with \texttt{top\_p}=1.0 and a maximum output length of 8192 tokens per invocation. Each task is allowed up to \textbf{100 ReAct iterations}. Parallel tool calls are disabled within a single agent so that browser operations are executed sequentially, whereas task-level parallelism is achieved through independent workers.

Before the official competition, we conducted local evaluations under Protocol~1. Each evaluation comprised 100 tasks executed continuously and sequentially using a single browser, with NavEval used to assess task success. The system achieved a best local pass rate of \textbf{79\%}. The official competition used the Protocol~3 task set and combined NavEval with manual verification. With \textbf{eight concurrent browser workers}, our system achieved a final pass rate of \textbf{59\%} and \textbf{ranked first overall}, winning the WebRetriever Challenge. Table~\ref{tab:results} summarizes these results.

\begin{table*}[t]
\centering
\caption{Best local evaluation result and official competition result.}
\label{tab:results}
\begin{tabular}{lllllc}
\toprule
\textbf{Stage} & \textbf{Protocol} & \textbf{Execution} & \textbf{Scale} & \textbf{Evaluation} & \textbf{Pass Rate} \\
\midrule
Local (best) & Protocol 1 & Single browser & 100 tasks & NavEval & 79\% \\
Official & Protocol 3 & 8 concurrent workers & 100 tasks & NavEval + manual & 59\% \\
\bottomrule
\end{tabular}
\end{table*}

Protocol~1 and Protocol~3 differ in task sets, execution settings, and evaluation procedures. Their scores therefore cannot directly quantify the effect of concurrent execution. These results characterize the complete system under the respective evaluation conditions; controlled ablations are needed to isolate the contributions of visual localization, context management, and fault recovery.

\section{Discussion}

The system design reflects a trade-off between semantic precision and visual coverage. Accessibility-tree references provide explicit interaction targets for standard controls, while screenshots reveal information that may be absent from structured representations. Coordinate localization extends the set of accessible controls but remains sensitive to page reflow and asynchronous updates. Keeping action execution with the main agent allows visual results to be interpreted alongside the task and current browser state. A useful next step is to select between structured and visual tools more adaptively, according to the information needed for the current decision.

Context management similarly balances compactness against preservation of task state. Retaining element references after snapshot truncation supports continued interaction, while recent messages preserve detailed feedback that a historical summary may omit. However, the current rules prioritize structural roles and repeated content rather than explicitly estimating task relevance. A short but decisive piece of evidence may therefore be lost during truncation or summarization. Recovery from summary failure keeps execution running but cannot ensure that all earlier findings remain available. Task-aware information selection and memory mechanisms that preserve supporting evidence are promising extensions.

Runtime recovery requires a distinction between completing the current task and preserving the ability to process subsequent tasks. Page-level errors can often be returned to the agent for another attempt, whereas connection loss may invalidate the current execution and require environment recovery. Bounded retries and worker restarts limit how long a failure occupies a browser, but do not by themselves recover an interrupted task. Future evaluation should examine both task success and execution continuity under different failure conditions, complementing the aggregate pass rates reported here.

\section{Conclusion}

This report presented our winning system for the WebRetriever Challenge. Built on AgentScope and Playwright MCP, it combines structured webpage interaction with screenshot understanding and grid-assisted visual localization. Hierarchical context management retains task progress across long trajectories, while browser-session reuse, bounded timeout handling, and worker-level recovery support concurrent execution. Together, these components address the interaction, memory, and runtime requirements of practical web-agent tasks.

The system achieved a best local pass rate of 79\% under Protocol~1 and ranked first in the official Protocol~3 competition with a 59\% pass rate using eight concurrent browser workers. Our experience emphasizes the need to coordinate model decision making with page representations and execution infrastructure; adaptive tool selection and task-aware memory offer directions for further improvement.


\begin{thebibliography}{99}

\bibitem{dong2026webretriever}
W. Dong, T. Fu, et al.,
``WebRetriever: A Large-Scale Comprehensive Benchmark for Efficient Web Agent Evaluation,''
arXiv preprint arXiv:2607.06118, 2026.
\url{https://arxiv.org/abs/2607.06118}

\bibitem{react}
S. Yao, J. Zhao, D. Yu, et al.,
``ReAct: Synergizing Reasoning and Acting in Language Models,''
ICLR, 2023.
\url{https://arxiv.org/abs/2210.03629}

\bibitem{agentscope}
AgentScope,
``AgentScope.'' Software repository.
\url{https://github.com/agentscope-ai/agentscope}

\bibitem{playwrightmcp}
Playwright MCP,
``Playwright MCP.'' Software repository.
\url{https://github.com/microsoft/playwright-mcp}

\bibitem{webvoyager}
H. He, W. Yao, K. Ma, et al.,
``WebVoyager: Building an End-to-End Web Agent with Large Multimodal Models,''
Proceedings of ACL, pp.~6864--6890, 2024.
\url{https://aclanthology.org/2024.acl-long.371/}

\end{thebibliography}
\end{document}